\documentclass[conference]{IEEEtran}
\IEEEoverridecommandlockouts
\usepackage{cite}
\usepackage{amsmath,amssymb,amsfonts}
\usepackage{algorithmic}
\usepackage{graphicx}
\usepackage{textcomp}
\usepackage{subcaption}     
\usepackage{xcolor}
\def\BibTeX{{\rm B\kern-.05em{\sc i\kern-.025em b}\kern-.08em
    T\kern-.1667em\lower.7ex\hbox{E}\kern-.125emX}}
\begin{document}
\title{A Novel Environment Object Modeling Method for Vehicular ISAC Scenarios\\
}
\author{\IEEEauthorblockN{
		\large Hanyuan Jiang$^*$,   % 1st author, 1st affiliations
		Yuxiang Zhang$^*$,   % 2nd author, 2nd affiliations
		Yameng Liu$^*$,    % 3rd author, 3rd affiliations
		Jianhua Zhang$^*$,
		Lei Tian$^*$,
		Tao Jiang$^\dagger$
	}
	\IEEEauthorblockA{%
		$^*$\emph{State Key Lab of Networking and Switching Technology, Beijing University of Posts and Telecommunications, Beijing, China}}
		$^\dagger$\emph{China Mobile Research Institution, Beijing, China}
	\IEEEauthorblockA{$^*$\{jianghy, zhangyx, liuym, tianlbupt, jhzhang\}@bupt.edu.cn}
	\IEEEauthorblockA{$^\dagger$jiangtao@chinamobile.com}
}

\maketitle

\begin{abstract}
Integrated Sensing and Communication (ISAC), as a fundamental technology of 6G, empowers Vehicle-to-Everything (V2X) systems with enhanced sensing capabilities. One of its promising applications is the reliance on constructed maps for vehicle positioning. Traditional positioning methods primarily rely on Line-of-Sight (LOS), but in urban vehicular scenarios, obstructions often result in predominantly Non-Line-of-Sight (NLOS) conditions. Existing researched indicate that NLOS paths, characterized by one-bounce reflection on building wall with determined delay and angle, can support sensing and positioning. However, experimental validation remains insufficient. To address this gap, channel measurements are conducted in an urban street to explore the existence of strong reflected paths in the presence of a vehicle target. The results show significant power contribution from NLOS paths, with large Environmental Objects (EOs) playing a key role in shaping NLOS propagation. Then, a novel model for EO reflection is proposed to extend the Geometry-Based Stochastic Model (GBSM) for ISAC channel standardization. Simulation results validate the model’s ability to capture EO's power and position characteristics, showing that higher EO-reflected power and closer distance to Rx reduce Delay Spread (DS), which is more favorable for positioning. This model provides theoretical guidance and empirical support for ISAC positioning algorithms and system design in vehicular scenarios.
\end{abstract}


Integrated Sensing and Communication (ISAC) is emerging as a fundamental technology to realize ubiquitous sensing and support digital twins within 6G networks \cite{liu2020vision}. Unlike conventional systems that are typically limited to either communication or sensing, ISAC combines both functions into a unified framework. This allows base stations and terminals to perform communication tasks while concurrently sensing the environment \cite{zhang2023channel}. Vehicle-to-Everything (V2X) serves as a typical application scenario for ISAC, playing a significant role in enabling autonomous driving\cite{cheng2022integrated,zhong2022empowering}. Notably, vehicle positioning is one of the key functions, relying on the acquisition of environmental information for map construction, and using algorithms to accurately determine the vehicle's location\cite{3GPP_22837}.

Traditional positioning algorithms, such as TOA (Time of Arrival) and TDOA (Time Difference of Arrival), are suitable for ideal LOS (Line of Sight) condition \cite{kuutti2018survey,guvenc2009survey}. However, in urban vehicular scenarios, the presence of obstacles such as buildings and trees adds complexity to the environment, resulting in significant multipaths effects that degrade positioning accuracy\cite{bader2022nlos}. To address this challenge, \cite{al2002ml} proposes using NLOS paths that undergo one-bounce reflection and leverages bidirectional estimation of the Angle of Arrival (AOA) and TOA. In addition, \cite{chen2024and} finds through ray-tracing simulations that NLOS positioning primarily relies on fixed and regular scatterers in the environment, such as walls. These scatterers are referred to as environment objects (EOs) while other multiple-bounce scattering paths should be discarded as much as possible.

However, most of these findings are based on theoretical analysis and simulation studies. They have not been validated by measurements to confirm whether urban vehicular scenarios contain EOs that can reflect one-bounce strong power NLOS paths for positioning. Furthermore, \cite{pyp} discusses the 3GPP's latest need for modeling EO to better characterize the propagation characteristics of ISAC channel in vehicular scenarios. Yet, this theory remains unsupported by experimental validation.

Therefore, this paper aims to address the aforementioned gaps by conducting bi-static measurements in an urban street environment, with a vehicle as sensing target at 26 GHz. The results reveal that the power of NLOS paths is sufficiently strong for sensing, primarily influenced by large EOs such as wall and metal fence. To quantify this impact, \(K_{EO}\)-factor is introduced to describe the proportional relationship between NLOS paths and EO-reflected paths. Furthermore, a novel EO reflection model is proposed based on ground reflection in Geometry-Based Stochastic Model (GBSM), offering a more accurate representation under NLOS condition in ISAC channel. Simulation results validate this model’s effectiveness in capturing both EO power and position  characteristics, as evidenced by shifts in the Cumulative Distribution Function (CDF) of Delay Spread (DS).

The remainder of the paper is organized as follows. Section II describes the measurement system and scheme. Section III presents the measurement results and conducts a quantitative analysis of the power contributions from different scatterers in the ISAC channel. Section IV focuses on the modeling of EO especially in NLOS condition, providing expressions based on ground reflection, along with the simulation results on DS under different power proportions and EO positions. Section V provides the conclusion of this study.

\section{MEASUREMENTS DESCRIPTION} 
\subsection{Measurement System}
\begin{figure}[h]
	\centering
	\includegraphics[width=0.4\textwidth]{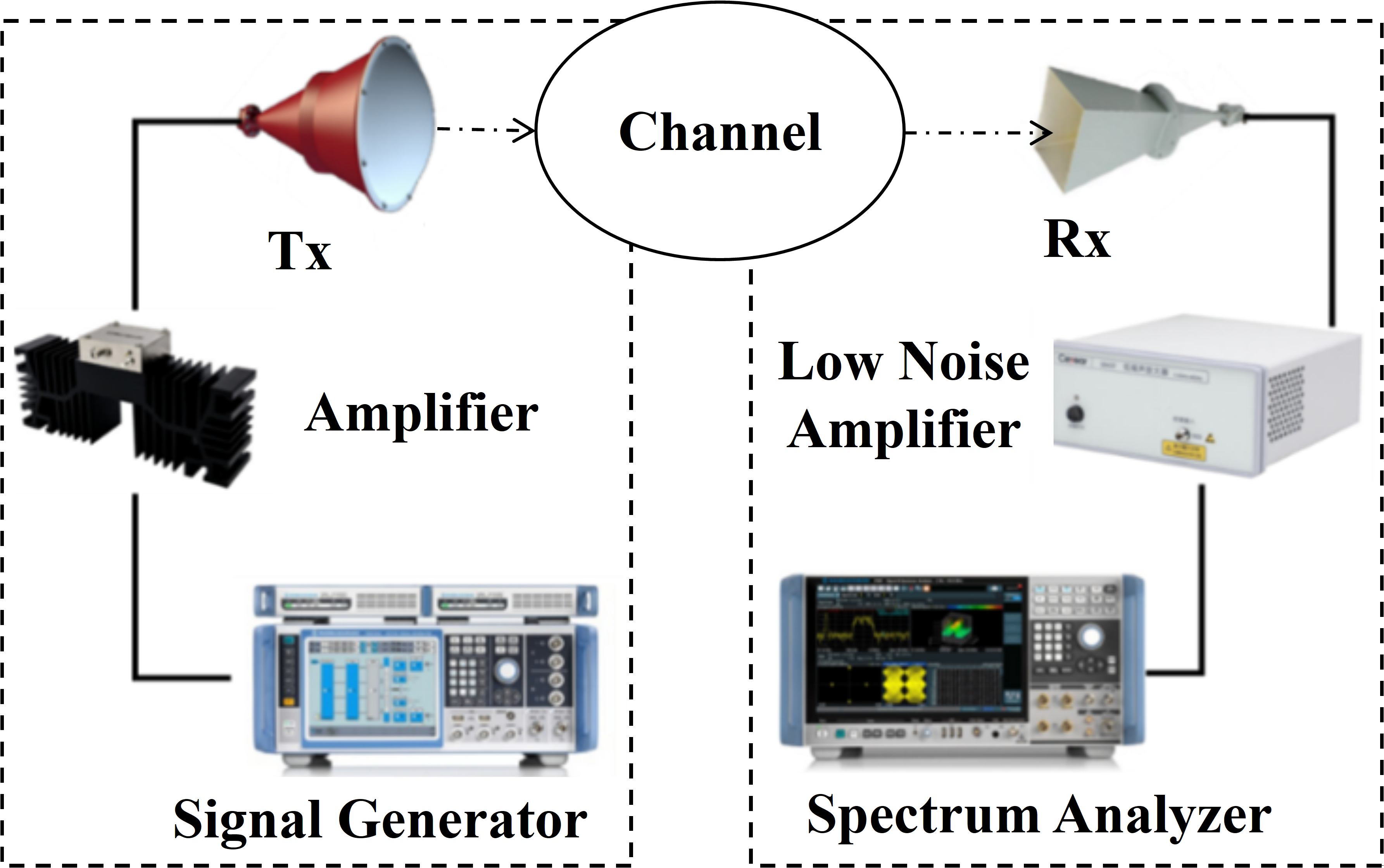} % 更改为你的图片文件名
	\caption{\centering Measurement system diagram.}
	\label{fig:2}
\end{figure}

In this measurement experiment, the Transmitter (Tx) utilizes a vector signal generator to generate a PN code sequence with a length of 1022 bits. The generated signal is then modulated to 26 GHz using binary phase shift keying and transmitted via a lens antenna, which provides excellent directivity and high gain, ensuring that more energy is directed toward the target. At the Receiver (Rx), a spectrum analyzer is used to capture and record the received signal. To ensure sufficient system gain, a low-noise amplifier is used at the receiver to amplify the signal\cite{zhang20173}. Fig.~\ref{fig:2} illustrates the structure of the measurement system. The specific configuration is detailed in Table.~\ref{tab:exp1}.
\begin{table}[h]
	\centering
	\caption{\centering Measurement System Configuration}

	\resizebox{0.9\columnwidth}{!}{
		\begin{tabular}{c|c} % @{} 用于去掉左右边距
			\hline
			\hline
			\textbf{Parameters} & \textbf{Value/Type} \\ 
			\hline
			Center frequency (GHz) & 26 \\
			\hline
			Bandwidth (MHz) & 600 \\
			\hline
			PN sequence & 1022 \\
			\hline
			Horn antenna azimuth HPBW (deg) & 8 \\
			\hline
			Lens Horn Antenna azimuth HPBW (deg) & 4.75 \\
			\hline
			Height of Tx/Rx Antenna (m) & 1.6 \\
			\hline
			Antenna polarization & V-to-V \\
			\hline
			\hline
		\end{tabular}
	}
	\label{tab:exp1}
\end{table}
\subsection{Measurement Scheme}
\begin{figure}[h]
	\centering
	\begin{subfigure}[b]{0.5\textwidth} % 设置子图宽度为0.6倍的行宽
		\centering
		\includegraphics[width=0.9\textwidth]{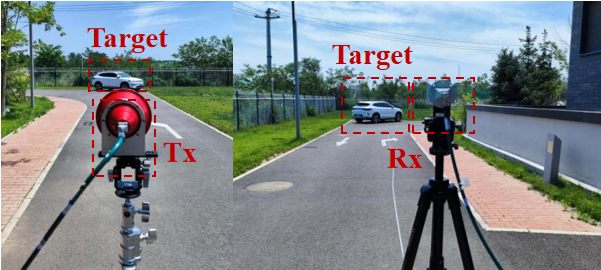} % 调整图像大小（宽度为子图的90%）
		\caption{Case1 scenario} % 图像的标题
		\label{fig:a} 
	\end{subfigure}
	\begin{subfigure}[b]{0.5\textwidth}
		\centering
		\includegraphics[width=0.9\textwidth]{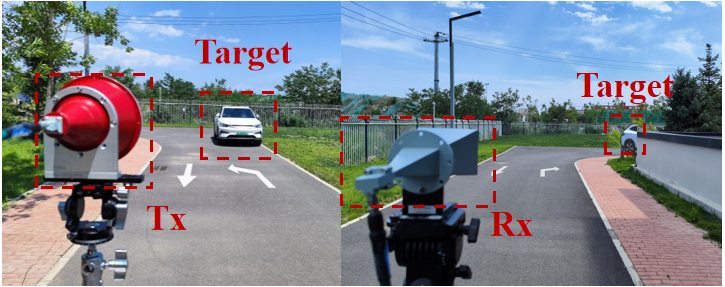} % 调整图像大小
		\caption{Case2 scenario}
		\label{fig:b}
	\end{subfigure}
	\caption{\centering The measurement campaign} % 整体图的标题
	\label{fig:two_figures}
\end{figure}

As depicted in Fig.~\ref{fig:two_figures}, a Bi-static ISAC channel measurement campaign is conducted along an urban L-shaped street. The selected environment facilitates clear scatterers identification including tall building wall, dense metal fence, lampposts, and vegetation. The Tx, equipped with a lens antenna, is positioned on the west side of the street transmitting eastward. A horn antenna on the north side faces south at a 0\textdegree\ angle. The horn antenna is rotated both clockwise and counterclockwise, collecting data every 5\textdegree\ across a range of ±90\textdegree, resulting in a total of 36 receiving angles. The absence of a LOS path between Tx and Rx is noted, obstructed by a tall building.

Two cases are designed to investigate the sensing of target by Rx in bi-static mode. Initially, a \(4.2m \times 1.7m \times 1.6m\) vehicle is positioned at a 45\textdegree\ angle in the corner of the street, facing the Tx (Case 1, Fig.\ref{fig:two_figures}(a)). In this scenario, the vehicle is aligned with the Tx, establishing a clear LOS path. This setup simulates a typical turning scenario, where the vehicle is in motion and about to change direction. In Case 2 (Fig.\ref{fig:two_figures}(b)), the vehicle is moved to the left side of the east-west street, facing west. In this situation, there is no direct alignment with the Tx, and there is also no clear LOS path to the Rx. This setup represents a scenario where the vehicle is about to complete its turn at the intersection.
\setlength{\textfloatsep}{5pt}  % 调整图片与正文的间距
%The measurement of environment, Case 1, and Case 2 involve constant positions for the Tx and Rx antennas with identical configuration and measurement methods. The only variation is the vehicle's placement. A top view schematic, shown in Fig.~\ref{fig:shce}, illustrates the setup, highlighting key scatterers and the size of the environment.

%\begin{figure}[h]
%	\centering
%	\includegraphics[width=0.4\textwidth]{shce.png} % 更改为你的图片文件名
%	\caption{\centering Schematic Diagram of of measurement scenario.}
%	\label{fig:shce}
%\end{figure}
\setlength{\textfloatsep}{5pt}  % 调整图片与正文的间距
\section{MEASUREMENT RESULTS AND ANALYSIS}
\begin{figure}[t]
	\centering
	\begin{subfigure}[b]{0.5\textwidth} % 设置子图宽度为0.6倍的行宽
		\centering
		\includegraphics[width=1\textwidth]{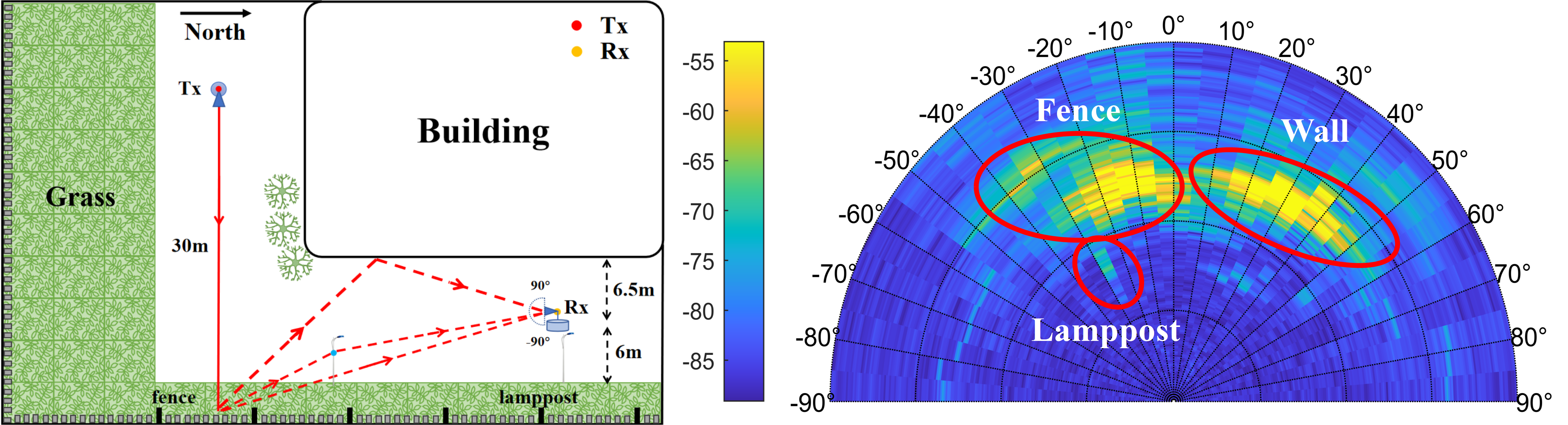} % 调整图像大小（宽度为子图的90%）
		\caption{}
		\label{fig:a} % 子图的标签
	\end{subfigure}
	\vspace{1em} % 在图之间添加垂直间距
	\begin{subfigure}[b]{0.5\textwidth} % 设置子图宽度为0.6倍的行宽
		\centering
		\includegraphics[width=1\textwidth]{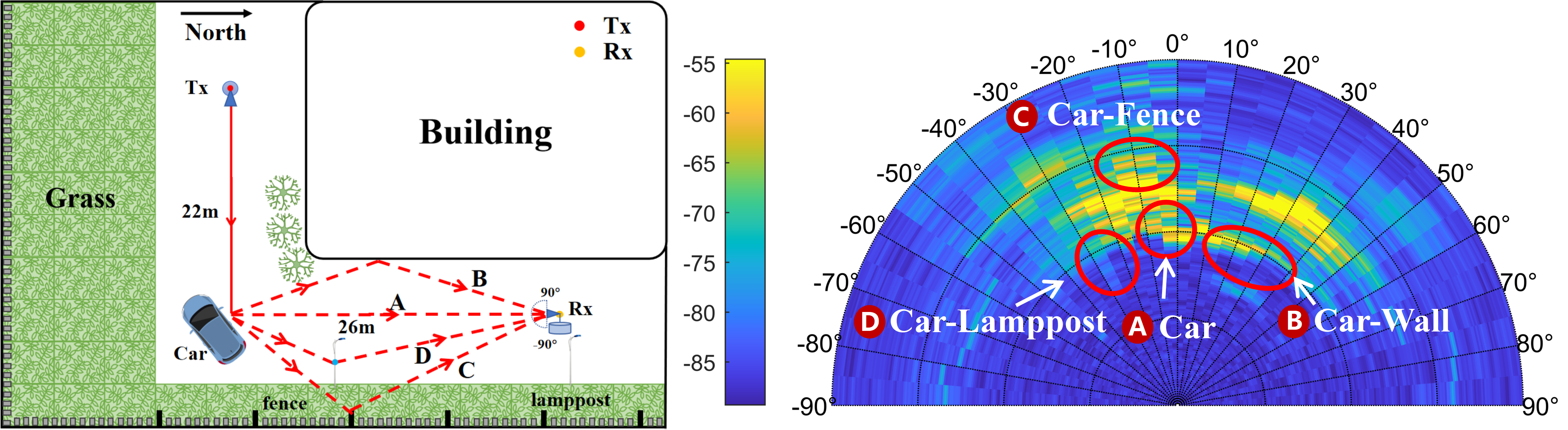} % 调整图像大小（宽度为子图的90%）
		\caption{}
		\label{fig:b} % 子图的标签
	\end{subfigure}
	\vspace{1em} % 在图之间添加垂直间距
	\begin{subfigure}[b]{0.5\textwidth}
		\centering
		\includegraphics[width=1\textwidth]{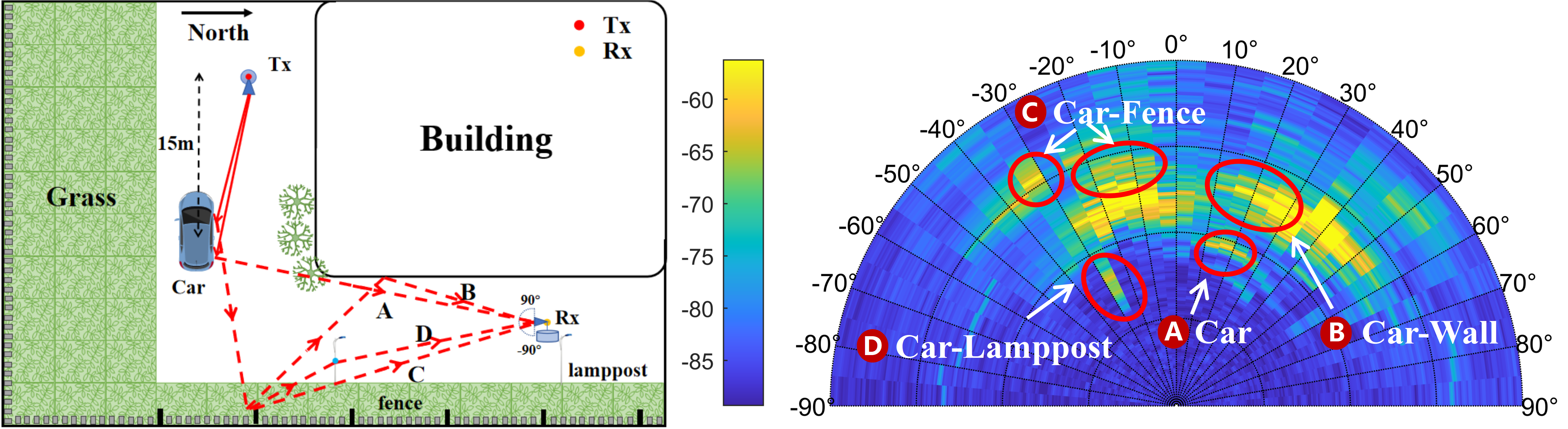} % 调整图像大小
		\caption{}
		\label{fig:c}
	\end{subfigure}
	\caption{\centering The analysis of possible paths and result of PADP for different cases are as follows: (a) Environment. (b) Case1. (c) Case2.} % 整体图的标题
	\label{fig:three_figures}
\end{figure}

To visualize the distribution of multipath components (MPCs) across delay, spatial, and power domains, Fig.~\ref{fig:three_figures} presents the Power-Angular Delay Profiles (PADPs) under various cases, along with the analysis of potential propagation paths. In the environment, the PADP clearly distinguishes various scatterers, such as the mental fence, wall, and lampposts, highlighting their individual contributions to the MPCs. When the vehicle is introduced into the environment, one-bounce reflection from the vehicle (label A) becomes clearly noticeable in Case 1. However, it is relatively weaker in Case 2 due to the absence of a distinct LOS path. By comparing the PADP of Case 1 and Case 2 with the baseline environment, the observed changes reveal the MPCs influenced by the vehicle and reflected from the surrounding scatterers, labeled as B, C and D. These MPCs that interact with the vehicle and are finally received can be defined as the target channel.

\begin{figure}[h]
	\centering
	\includegraphics[width=0.5\textwidth]{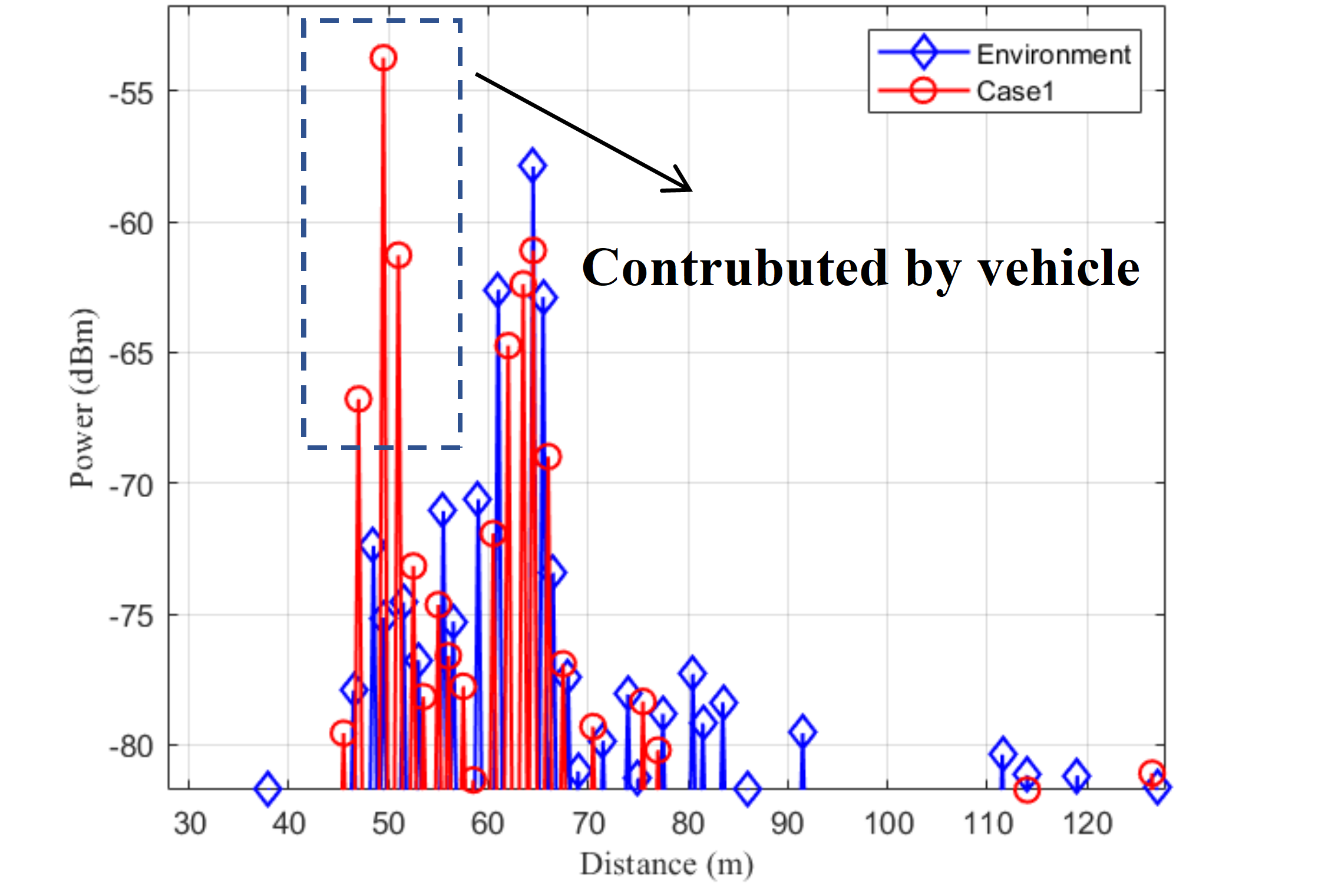} % 更改为你的图片文件名
	\caption{\centering Illustration of the comparison of MPCs at Rx with 5° angle between the environment and Case 1. The additional red lines at the 48m position indicate the effective MPCs contributed by the vehicle target.}
	%	\label{fig:duibi}
	\label{fig:duibi}
\end{figure}

Subsequently, a quantitative analysis on the power proportion of different paths in target channel is conducted to further investigate which scatterers matter in the vehicular scenario. First, we accurately measure the size of the environment, calculating the distances and angle ranges that the signal propagates from Tx through various sactterers to the receiver, ensuring the precise identification of the corresponding MPCs. By comparing the Power Delay Profiles (PDP) of the environment, Case 1, and Case 2 at the same angles, MPCs attributed by the vehicle are identified as shown in Fig.~\ref{fig:duibi}. Moreover, measurement data reveal that wall and fence, as relatively regular scatterers, exhibit strong energy at receiver angles when facing them at specific incident and reflection angles, which suggests the adherence to specular reflection mechanism. Therefore, the MPCs contributed by the target and then reflected by different scatterers can be separated at all Rx angles.

 Then, the power contribution \( PP_{scatterer} \) of each scatterer to the target channel is calculated, as shown in (\ref{equ:4}). In this equation, \( P_{scatterer} \) represents the total power from a specific scatterer across \(N\) measured angles, which is 36 in this measurement. While \( P_{tar} \) denotes the total power of the target channel across all angles, which includes the sum of power contributions from all scatterers. This method quantifies the relative contribution of each scatterer to the overall target channel power, helping to identify which scatterers have the most significant impact. Table.\ref{tab:case1} and Table.\ref{tab:case2} summarize the power proportions of different paths in target channel.
\begin{equation}
	PP_{scatterer} = \frac{P_{scatterer}}{P_{tar}}=\frac{\displaystyle\sum_{i=1}^{N} P_{scatterer, i}}{\displaystyle\sum_{i=1}^{N} P_{tar, i}}
	\label{equ:4}
\end{equation}
\begin{figure*}[t]  % 使用 figure* 环境，并指定 [b] 强制将公式放在页面底部
	% \hrulefill  % 在公式上方绘制一条横线
	\centering  % 让公式居中显示
	\begin{align}
		H_{u,s}^{\text{NLOS}}(\tau,t) = \sqrt{K_{EO}} \cdot \sum_{k=1}^{K} H_{u,s,k}^{\text{EO}}(t) \delta(\tau - \tau_{EO,k}) + \sqrt{1 - K_{EO}} \cdot \sum_{n=1}^{N} H_{u,s,n}^{\text{NLOS}}(t) \delta(\tau - \tau_n)\tag{3}
		\label{equ:h}
	\end{align}
\end{figure*}

\begin{figure*}[t]  % 使用 figure* 环境，并指定 [b] 强制将公式放在页面底部
	% \hrulefill  % 在公式上方绘制一条横线
	\centering  % 让公式居中显示
	\begin{align}
		H_{u,s}^{\text{EO}}(t, \tau) &= 
		\left[
		\begin{array}{c}
			F_{rx,u,\theta}\left( \theta_{EO,ZOA}, \phi_{EO,AOA} \right) \\
			F_{rx,u,\phi}\left( \theta_{EO,ZOA}, \phi_{EO,AOA} \right)
		\end{array}
		\right]^{\text{T}}
		\cdot
		\left[
		\begin{array}{cc}
			R^{\text{EO}}_{\parallel} & 0 \\
			0 & -R^{\text{EO}}_{\perp}
		\end{array}
		\right]
		\cdot
		\left[
		\begin{array}{c}
			F_{tx,s,\theta}\left( \theta_{EO,ZOD}, \phi_{EO,AOD} \right) \\
			F_{tx,s,\phi}\left( \theta_{EO,ZOD}, \phi_{EO,AOD} \right)
		\end{array}
		\right] \notag \\ 
		& \quad \cdot \exp\left( -j 2 \pi \frac{d_{EO}}{\lambda_0} \right) 
		\exp\left( j 2 \pi \frac{\hat{r}_{rx,EO}^T \cdot \mathbf{d}_{rx,u}}{\lambda_0} \right) 
		\cdot \exp\left( j 2 \pi \frac{\hat{r}_{tx,EO}^T \cdot \mathbf{d}_{tx,s}}{\lambda_0}\right) \notag \\
		& \quad \cdot \exp\left( j 2 \pi \frac{\hat{r}_{tx,EO}^T \cdot \mathbf{v}_{tx}}{\lambda_0}t\right)
		\cdot \exp\left( j 2 \pi \frac{\hat{r}_{rx,EO}^T \cdot \mathbf{v}_{rx}}{\lambda_0}t\right) \tag{4} 
		\label{equ:eo}
	\end{align}
\end{figure*}

\begin{figure*}[t]  % 使用 figure* 环境，并指定 [b] 强制将公式放在页面底部
	% \hrulefill  % 在公式上方绘制一条横线
	\centering  % 让公式居中显示
	\begin{align}
		\tau_{EO} = \frac{d_{EO}}{c} = \frac{\sqrt{(h_{tx} - h_{rx})^2 + (d_{tx} + d_{rx})^2 + d_{2D}^2-(d_{tx} - d_{rx})^2}}{c}\tag{5}
		\label{equ:tau_eo}
	\end{align}
\end{figure*}

\renewcommand{\arraystretch}{1.5} % 调整行间距
\begin{table}[h]
	\centering
	\caption{\centering Power proportion of different paths in target channel for Case 1}
	\Large 
	\resizebox{1\columnwidth}{!}{
		\begin{tabular}{c|c|c|c|c} % 添加一个 'c' 来匹配列数
			\hline
			\hline
			Path & Contributed by & Propagation & Distance/Angle & Power proportion \\ 
			\hline
			A & Vehicle & LOS-LOS & 47m$\sim$51m (-5°$\sim$ 5°) & 50.5\%\\
			\hline
			B & Vehicle-Wall & LOS-NLOS & 50m$\sim$53m (15°$\sim$55°) & 9.7\%\\
			\hline
			C & Vehicle-Fence & LOS-NLOS & 52m$\sim$58m (-55°$\sim$-15°) & 34.5\%\\
			\hline
			D & Vehicle-Lamppost & LOS-NLOS & 49m$\sim$50m (-30°$\sim$-25°) & 5.3\%\\
			\hline
			\hline
		\end{tabular}
	}
	\label{tab:case1}
\end{table}
\begin{table}[h]
	\centering
	\caption{\centering Power proportion of different paths in target channel for Case 2}
	\Large 
	\resizebox{1\columnwidth}{!}{
		\begin{tabular}{c|c|c|c|c} % 添加一个 'c' 来匹配列数
			\hline
			\hline
			Path & Contributed by & Propagation & Distance/Angle & Power proportion \\ 
			\hline
			A & Vehicle & LOS-LOS & 45m$\sim$48m (-5°$\sim$5°) & 17.9\%\\
			\hline
			B & Vehicle-Fence-Wall & LOS-NLOS & 60m$\sim$66m (15°$\sim$55°) & 61.3\%\\
			\hline
			C & Vehicle-Fence & LOS-NLOS & 52m$\sim$58m (-55°$\sim$-5°) & 8.4\%\\
			\hline
			D & Vehicle-Fence-lamppost & LOS-NLOS & 49m$\sim$50m (-30°$\sim$-25°) &12.4\%\\
			\hline
			\hline
		\end{tabular}
	}
	\label{tab:case2}
\end{table}
The measurement results indicate that when a vehicle target is introduced, environmental scatterers impact the target channel to varying degrees. Small scatterers like lampposts contribute relatively little power, around 12\% in Case 2 and only 5\% in Case 1. In contrast, large regular scatterers such as wall and metal fence significantly affect the target channel. In Case 1, besides direct reflection from vehicle, the metal fence and wall contribute about 45\% of the power, while in Case 2, due to the absence of LOS paths, they contribute approximately up to 70\%. These scatterers can be defined as EOs, which play a critical role in the target channel. As a result, the power from EOs is sufficiently strong to be leveraged for NLOS-assisted sensing.

\section{Modeling of Environment Object} 
\subsection{EO modeling in GBSM}
\begin{figure}[h]
	\centering
	\includegraphics[width=0.4\textwidth]{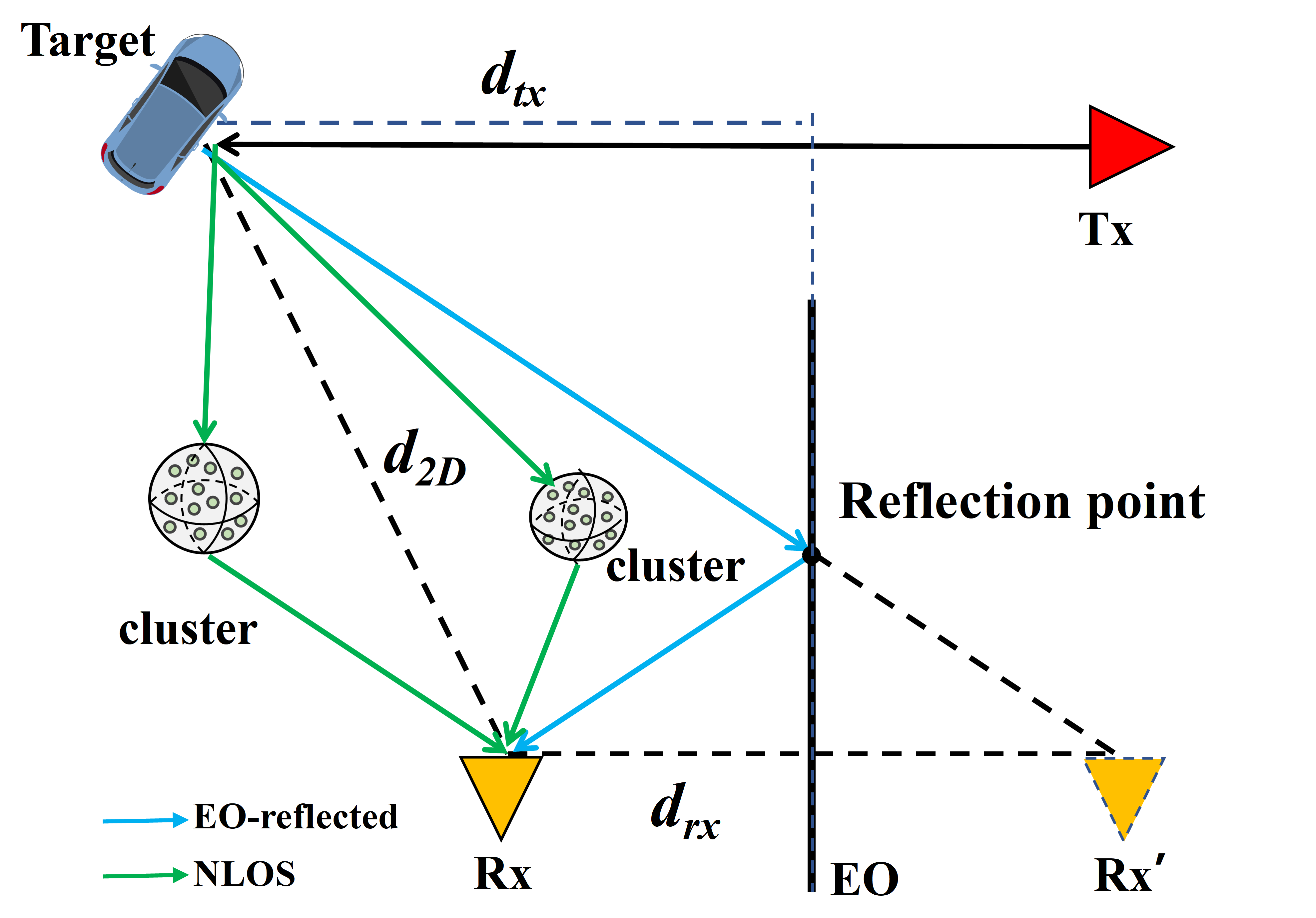} % 更改为你的图片文件名
	\caption{\centering Illusion of EO reflection.}
	\label{fig:eo}
\end{figure}

The previous section reveals that, in urban vehicular ISAC scenarios, a significant portion of the power in target channel originates from EOs, making it essential to model these components for positioning. In the bi-static sensing mode, due to the presence of target, the channel is divided into Tx-target and target-Rx components, which exhibit the same characteristics due to channel reciprocity, as illustrated in Fig.\ref{fig:eo}. Therefore, This section focuses on modeling the EO within target-Rx channel, assuming the vehicle acts as a transmitter in target channel. What's more, similar to the ground reflection model in GBSM\cite{3GPP_38901}, EOs can be treated as specular reflection surfaces and are typically considered strong paths under NLOS conditions, as indicated by the previous measurement results. 

To quantify the power contribution of the EO-reflected path, the \(K_{EO}\)-factor is proposed, as defined in (\ref{equ:r}). Here, \(P_{EO}\) and \(P_{NLOS}\) denote the total power of EO-reflected paths and the overall NLOS power in the environment, respectively. This parameter serves as a measure of the presence and influence of EOs within the environment. Specifically, when \(K_{EO} = 0\), it indicates the absence of significant EOs, whereas a higher \(K_{EO}\) value suggests that the majority of NLOS power originates from EO-reflected paths.

\begin{equation}
	K_{EO} = \frac{P_{EO}}{P_{NLOS}}, \quad K_{EO} \in [0, 1]\tag{2}
	\label{equ:r}
\end{equation}

When considering EO-reflected path in NLOS propagation, the total channel impulse response (CIR) under NLOS condition is expressed in (\ref{equ:h}). The total NLOS channel response \(H_{u,s}^{\text{NLOS}}\)is the summation of the EO-reflected paths and other NLOS paths, where \(K\) and \(N\) represent the total number of EO-reflected paths and other NLOS paths, respectively. The CIR for the original NLOS paths \(H_{u,s,n}^{\text{NLOS}}\) is generated based on the steps described in \cite{3GPP_38901}. For EO-reflected path within a 3D-MIMO system\cite{zhang2014three}, the CIR is defined by (\ref{equ:eo}), where
\begin{itemize}
	\item \( F_{rx, u, \theta} \) and \( F_{rx, u, \phi} \) are the field patterns of receive antenna element \( u \) the direction of the spherical basis vectors, \( \hat{\theta} \) and \( \hat{\phi} \) respectively.
	\item \( F_{tx, s, \theta} \) and \( F_{tx, s, \phi} \) are the field patterns of transmit antenna element \( s \) in the direction of the spherical basis vectors, \( \hat{\theta} \) and \( \hat{\phi} \) respectively. 
	\item \( \theta_{EO,ZOA} \) and \( \phi_{EO,AOA} \) are the zenith and azimuth angle of arrival after interacting with the EO.
	\item \( \theta_{EO,ZOD} \) and \( \phi_{EO,AOD} \) are the zenith and azimuth angle of departure after interacting with the EO.
	\item \(R_{\parallel}^{\text{EO}} and R_{\perp}^{\text{EO}}\) represent the reflection coefficients for parallel and perpendicular polarization, which vary with different materials and surfaces of EO.
	\item \(d_{EO}\) represents the propagation distance of EO-reflected path.
	\item \(\lambda_0\) is the wavelength of the carrier frequency.
	\item \(\hat{r}_{tx,EO}, \hat{r}_{rx,EO}\) indicate the unit vectors from the Tx or Rx to the EO, respectively.
	\item \(\mathbf{d}_{tx,u}, \mathbf{d}_{rx,s}\) indicate the location vectors of the transmitter and receiver antennas.
	\item \(\mathbf{v}_{tx}, \mathbf{v}_{rx}\) indicate the velocity vectors of transmitter and receiver.
\end{itemize}

Since NLOS-assisted positioning primarily relies on one-bounce reflected paths, while multiple-bounce paths are treated as noise that affect positioning accuracy, the EO modeling above mainly focuses on one-bounce reflections. Therefore, the position of EO can be determined by defining the horizontal distances \(d_{tx}\) and \(d_{rx}\) from the Tx and Rx to EO, respectively. Consequently, the delay, angles, and other parameters can be derived based on geometric relationships and theoretical formula, as shown in (\ref{equ:tau_eo}), where \(h_{tx}\) and \(h_{rx}\) are the height of Tx and Rx respectively. \(d_{2D}\) indicates the horizon distance between Tx and Rx.

\subsection{Numerical Analysis}
To validate the impact of EO modeling on the channel, we conduct simulations of real-world scenarios and compare the results with measurement data. Based on the formulas, the target-Rx channel in Urban Microcell(Umi) scenario with a single EO as reflecting wall surface can be simulated. The configuration parameters are shown in Table \ref{tab:sim}, with \( d_{tx} = d_{rx} = 6.5 \, \text{m} \) and \( K_{EO} = 0.5 \), which is the same deployment as the measurement described in section II. 
\begin{table}[h]
	\centering
	\caption{Simulation Configuration}
	\resizebox{0.7\columnwidth}{!}{
		\begin{tabular}{c|c} % @{} 用于去掉左右边距
			\hline
			\hline
			\textbf{Parameters} & \textbf{Value/Type} \\ 
			\hline
			Scenario & Umi urban street\\
			\hline
			Link condition & NLOS \\
			\hline
			Center frequency (GHz) & 26 \\
			\hline
			Bandwidth (MHz) & 600 \\
			\hline
			Number of Tx/Rx & 1 \\
			\hline
			Coordinates of Tx/Rx (m) & (0,0,1.6)/(0,26,1.6) \\
			\hline
			Type of Tx/Rx antenna & Horn \\
			\hline
			\hline
		\end{tabular}
	}
	\label{tab:sim}
\end{table}

\begin{figure}[h]
	\centering
	\begin{subfigure}[b]{0.5\textwidth} % 设置子图宽度为0.6倍的行宽
		\centering
		\includegraphics[width=0.9\textwidth]{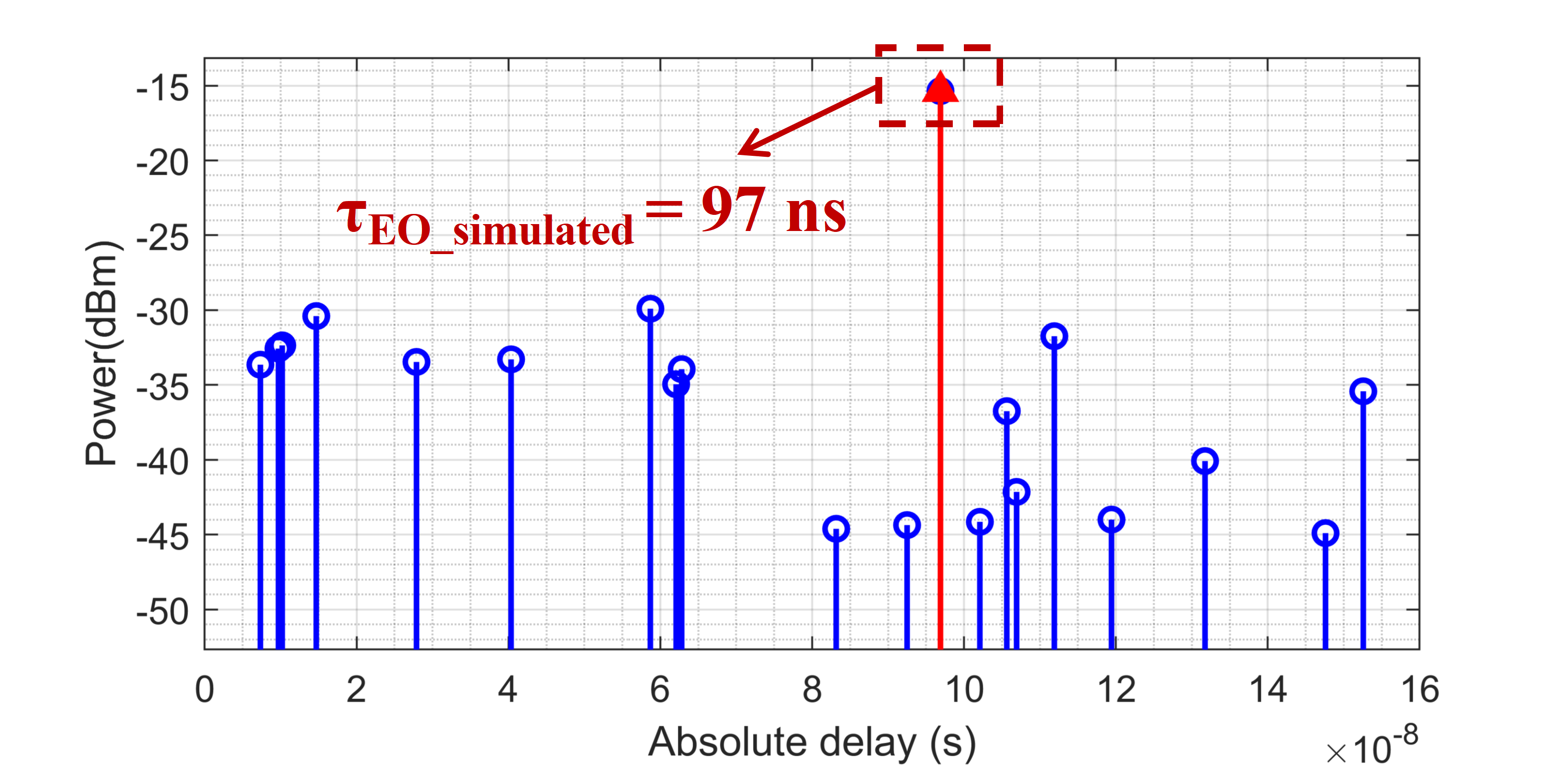} % 调整图像大小（宽度为子图的90%）
		\caption{} % 图像的标题
		\label{fig:A} 
	\end{subfigure}
	\begin{subfigure}[b]{0.5\textwidth}
		\centering
		\includegraphics[width=0.9\textwidth]{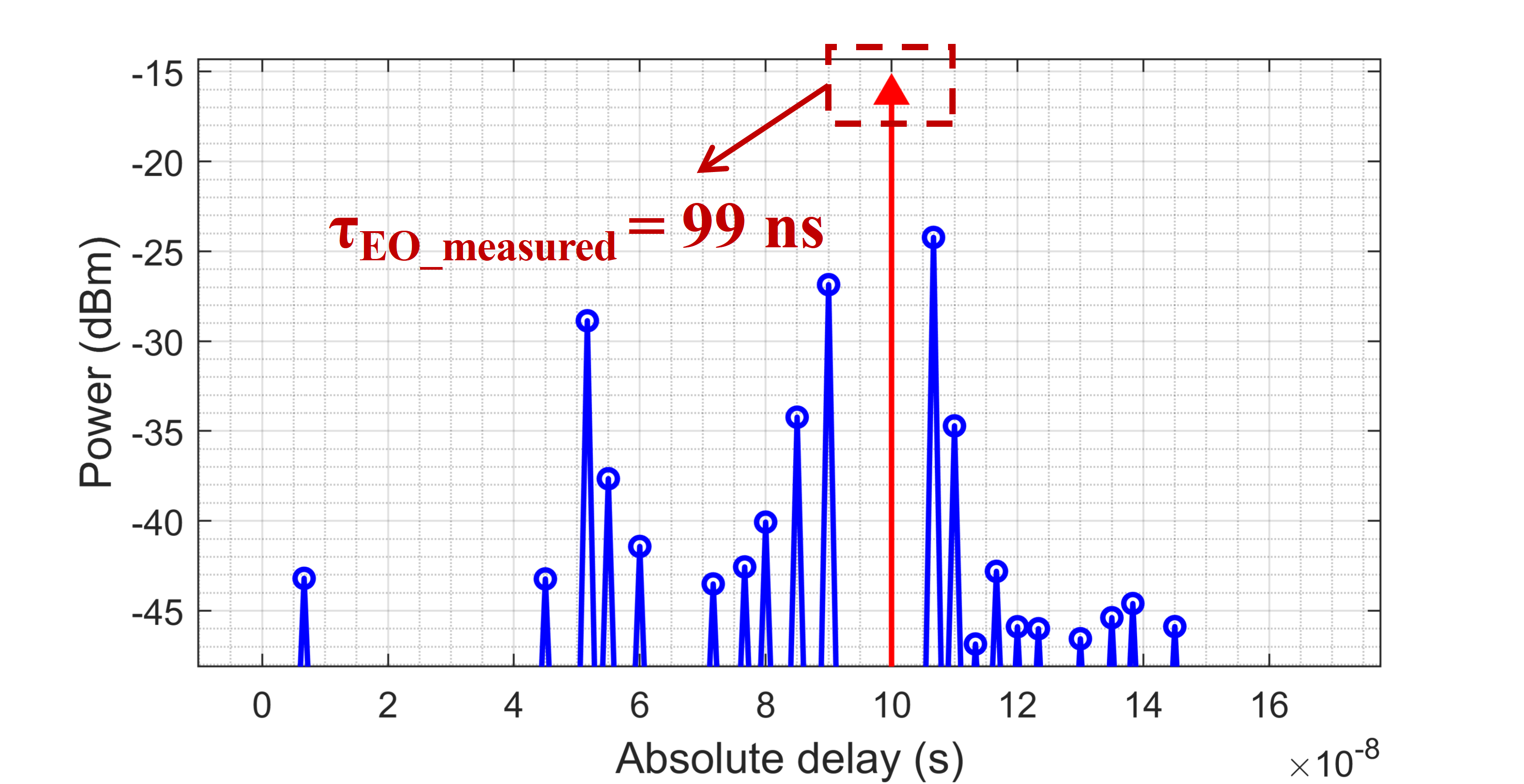} % 调整图像大小
		\caption{}
		\label{fig:B}
	\end{subfigure}
	\caption{\centering Simulated PDP result (a) is based on the configuration parameters in Table\ref{tab:sim}, which is corresponding to the measurement result (b).} % 整体图的标题
	\label{fig:pdp}
\end{figure}

Following simulation configuration described above, the PDP of target-Rx channel is plotted, as shown in Fig.~\ref{fig:pdp}. It can be observed that a distinct strong path appears at delay of 97 ns, which closely aligns with the absolute delay of 99 ns for the path reflected from target to the wall and then to Rx in the actual measurement scenario. The remaining MPCs describe the NLOS paths that reflected from other surrounding scatterers in the environment.
%\begin{figure}[h]
%	\centering
%	\includegraphics[width=0.5\textwidth]{PDP.png}\label{fig:left figure}
%	\caption{\centering Simulated PDP results based on the configuration parameters in Table\ref{tab:sim}, which is corresponding to the measurement deployment.}
%	\label{fig:pdp}
%\end{figure}
\begin{figure}[h]
	\centering
	\includegraphics[width=0.5\textwidth]{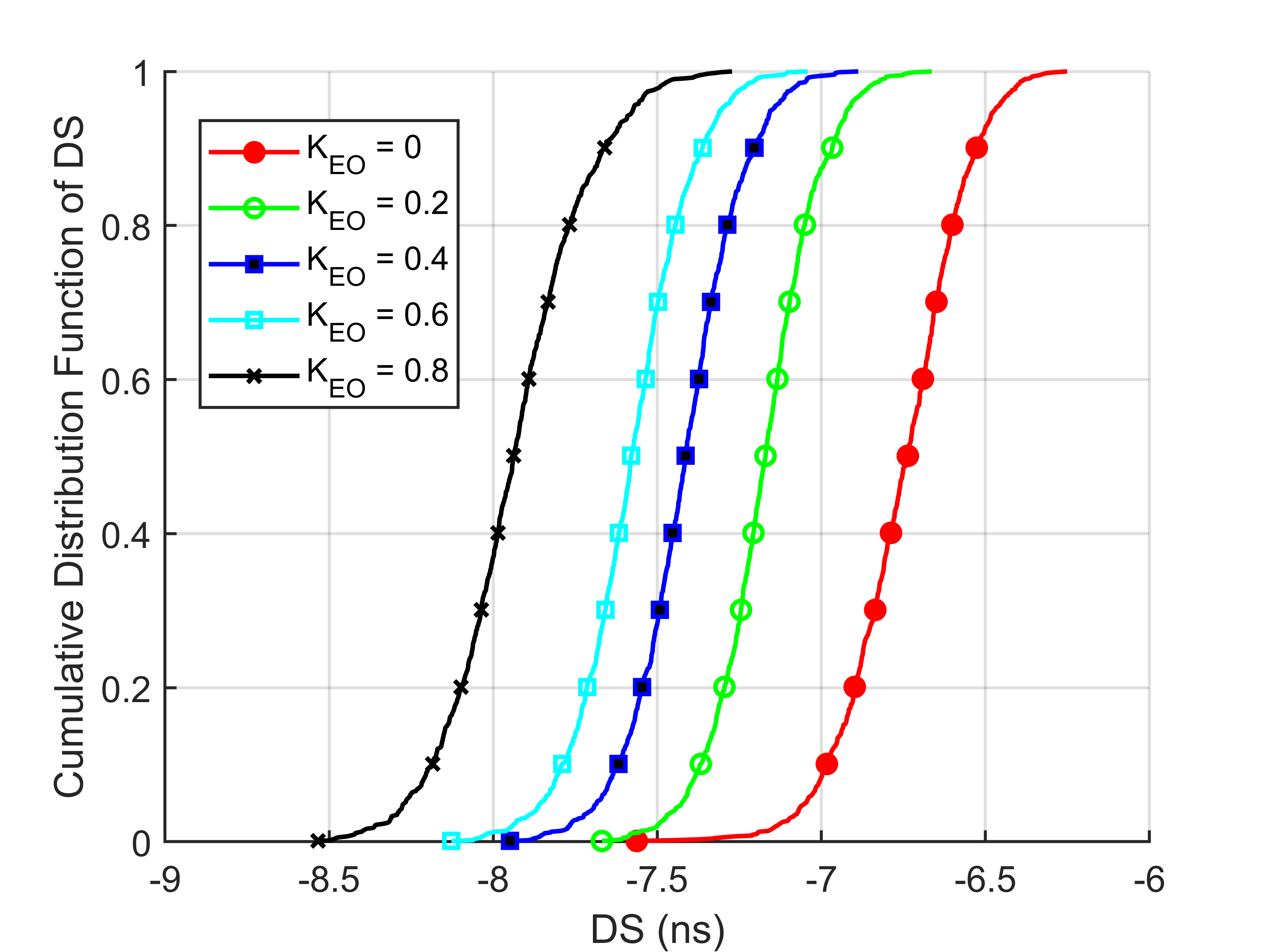} % 更改为你的图片文件名
	\caption{\centering CDF of DS under different \(K_{EO}\) values.}
	\label{fig:rcomp}
\end{figure}
\begin{figure}[h]
	\centering
	\includegraphics[width=0.5\textwidth]{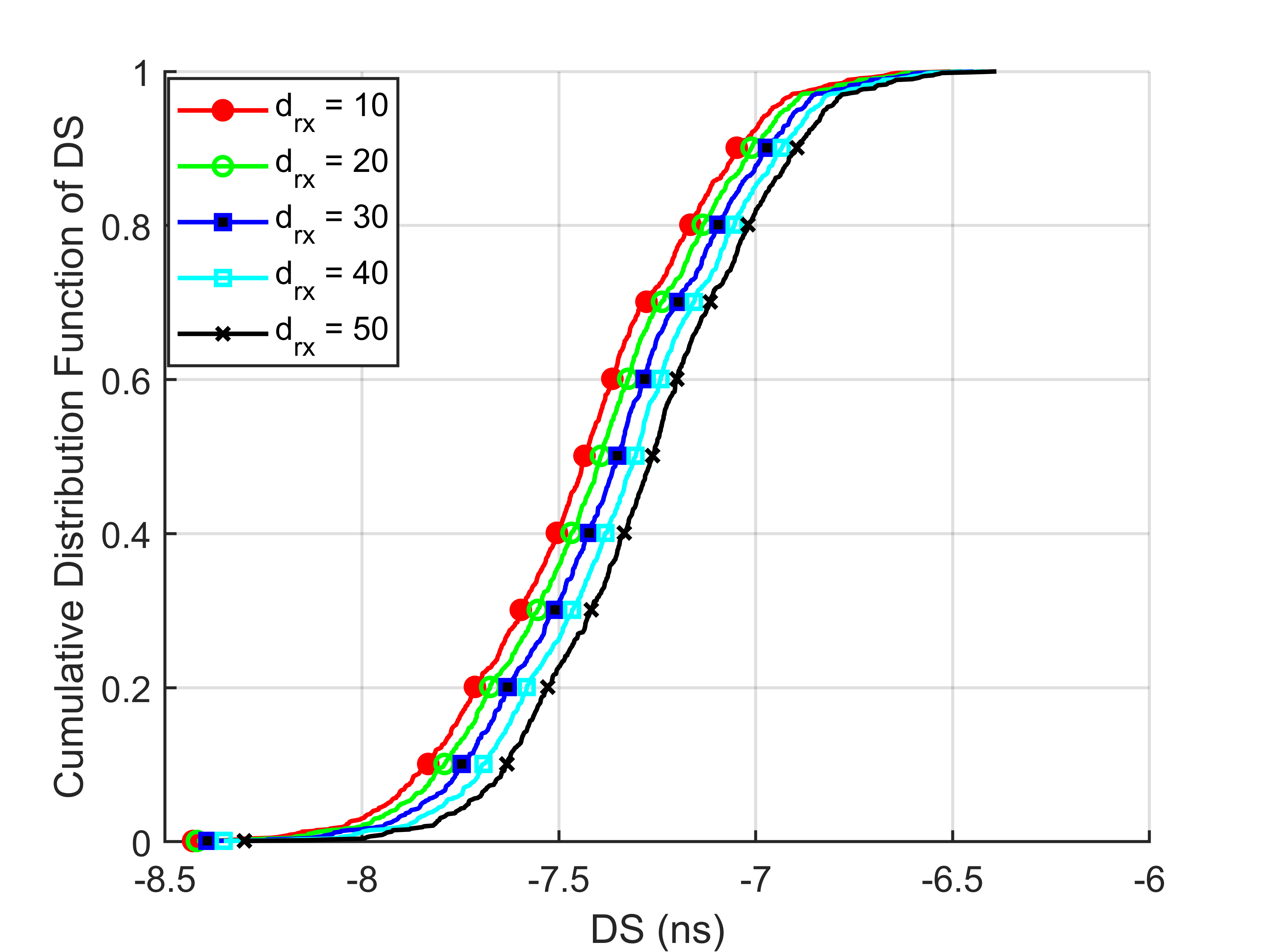} % 更改为你的图片文件名
	\caption{\centering CDF of DS under different \(d_{rx}\) values.}
	\label{fig:dcomp}
\end{figure}

Since the delay spread (DS) reflects the multipath effects of the channel \cite{jiang20213gpp}, a smaller DS indicates more concentrated energy, which is beneficial for improving positioning accuracy. Therefore, to further investigate the impact of introducing an EO on channel delay characteristics, simulations are conducted. Firstly, the \(K_{EO}\)-factor is varied to simulate different power contributions from the EO-reflected path, enabling the observation of changes in the CDF of DS under varying reflection intensities. Furthermore, the power contribution in the NLOS environment is kept constant, while the distance between the EO and Rx is adjusted to examine how the position of EOs affects the CDF of DS. 

The simulation results, as shown in Fig.~\ref{fig:rcomp} and ~\ref{fig:dcomp}, demonstrate that when the EO remains in the fixed position but the reflection power increases, the average DS decreases. This suggests that as more energy is focused on the EO-reflected path, the impact of other MPCs become smaller, which is more favorable for target sensing and positioning. Conversely, when the distance between the EO and the transmitter increases while keeping the power contribution constant, the average DS increases. This implies that as the EO moves farther from the Rx, its influence on the channel weakens, leading to a more dispersed signal and indicating a lack of EOs in the environment that can assist in positioning. These findings highlight the critical role of the EO's reflection strength and location in influencing the target's sensing and positioning capabilities within the ISAC channel.

\section{Conclusion}
This paper presents channel measurements in an urban vehicular scenario, emphasizing the notable power contributions reflected from large regular EOs in the target channel, which are favorable for sensing and positioning. An EO reflection model based on ground reflection in GBSM is introduced and validated through simulations, effectively capturing both EO power and position characteristics. These findings provide valuable insights for ISAC channel modeling and simulation in vehicular scenarios. Future research should focus on developing advanced algorithms and model designs to better leverage EO characteristics, further enhancing positioning and sensing performance.
\section*{Acknowledgment}
This research is supported by National Key R\&D Program of China (2023YFB2904802), Young Scientists Fund of the National Natural Science Foundation of China (62201087), The National Natural Science Foundation of China (U21B2014), Guangdong Major Project of Basic and Applied Basic Research (2023B0303000001), Key Program of Beijing Municipal Natural Science Foundation (L243002), and Beijing University of Posts and Telecommunications-China Mobile Research Institute Joint Innovation Center.
\bibliographystyle{IEEEtran}  % 或其他样式，如 plain, alpha 等
\bibliography{ref}  % 你的 .bib 文件的名字，不用扩展名

\end{document}